# Multiband Vector Plasmonic Lattice Solitons

Yao Kou, Fangwei Ye,* and Xianfeng Chen

*State Key Laboratory of Advanced Optical Communication Systems and Networks, Department of Physics, Shanghai Jiao Tong University, Shanghai 200240, China*
*\*Corresponding author: fangweiye@sjtu.edu.cn*

We predict multiband vector Plasmonic Lattice Solitons (PLSs) in metal-dielectric waveguide arrays, in both focusing and defocusing nonlinearities. Such vector solitons consist of two components originating from different transmission bands. By simulating the full nonlinear Maxwell's equations (MEs), we demonstrate the diffractionless propagation of vector PLSs and their discrete diffraction when only one component is present. Their subwavelength size characteristics and the influences of metallic losses are also studied. © 2010 Optical Society of America

Spatial vector solitons form when a delicate balance between diffraction, self-, and cross-modulation in all components is achieved. The components in a vector soliton may differ in frequencies, polarizations or they could belong to different spatial modes of their jointly induced trapping potentials [1-4]. In optically periodic systems, components can even spectrally locate in different bandgaps of the transmission spectrum [5-7]. Due to their fundamental physical interests and potential applications in all-optical technology, vector solitons have been studied in various structures [2-7].

Recently, there is an increasing interest in the study of nonlinear spatial optics in metallic nano-structures [8-17], where the surface plasmon polaritons (SPPs) at metal-dielectric interfaces greatly enhance the field amplitude and provide strong nonlinear effects under low power excitations. This facilitates the formation of plasmonic solitons [8]. One important class of plasmonic solitons is the so-called *plasmonic lattice solitons (PLSs)* [11-15] supported by nonlinear plasmonic lattices, i.e., a nanostructure composed of alternative metallic and nonlinear dielectric materials. When SPP tunneling in plasmonic lattices is inhibited by nonlinearity, PLSs form. A novel feature of PLSs is their possible deep-subwavelength spatial confinement which may find applications in nonlinearity-controlled nanodevices. However, to provide a possibility of light-controlling-light scheme at nanoscales, it is desirable to study the coherent and incoherent interactions between plasmonic solitons. A striking outcome of the nonlinear incoherent interactions of two PLSs might be the vector PLSs, which is the aim of the present study.

In this Letter, we reveal the existences of vector PLSs in nonlinear plasmonic arrays whose two components originate from Bloch modes of different bands of the transmission spectrum. The mutual incoherence of the two components is achieved by their widely-separated frequencies. As plasmonic structures are necessary highly dispersive, two components "see" different trapping potentials. Still, they mutually trap each other and form a composite state. The stationary and stable propagation of vector PLSs are demonstrated by a direct propagation simulation of the full nonlinear Maxwell's equations (MEs). The influence of metallic losses on soliton propagation is also addressed, and vector PLSs in self-defocusing nonlinearity are found of much longer propagation distances than those in self-focusing nonlinearity. The reason behind it is elucidated.

The geometry we consider is a 1D plasmonic lattice with periodically stacked metal and nonlinear dielectric layers along the x axis. Field propagation is assumed along z axis. The thickness of metal and dielectric layers is denoted by $t_m$ and $t_d$, respectively, which will be properly set for the emergence of higher-order transmission band. The intensity-dependent refractive index of the nonlinear layers is assumed as: $n_{NL} = \sqrt{\varepsilon_{NL}} = 3.5 + n_2 |E|^2$, where $n_2 = \pm 1.5 \times 10^{-19} m^2/V^2$ is the self-focusing/self-defocusing nonlinear coefficient. The complex permittivity of metal, $\varepsilon_m$, strongly depends on the wavelength [18]. The imaginary part of $\varepsilon_m$ is taken into account in the following, unless stated otherwise. In order to decrease the ohmic loss, we focus on the solitons in the infrared region.

The transmission bands of the plasmonic lattice is obtained by solving linear MEs for TM waves ($E_y=H_x=H_z=0$), where the nontrivial field components are expressed by $[E_x, H_y] = [A(x), B(x)]e^{i(\beta z - \omega t)}$ with $[A(x+T), B(x+T)] = [A(x), B(x)]e^{ik_x T}$ being Bloch mode profiles and $T = t_m + t_d$ the lattice periodicity. Figure 1(a) shows the band spectrum for $t_m$=50 nm and $t_d$=240 nm, at two wavelengths, $\lambda_{1(2)}$=1550(1310) nm. A gap opens for both wavelengths. Note that although the bandgap structures depicted here are found with the metallic dissipation into consideration, our comparison with the lossless cases reveals that the influence of dissipation on bandgap is negligible(both in shifting the gap edges and in changing the gap width)[19]. This is connected to the fact that the imaginary parts of the metallic permittivity is very small with respect to their real counterparts($\varepsilon_m$=-129+3.28i@1550nm, -92+2.13i@1310nm). Figure 1(b) studies the evolutions of band structures with the thickness of metal layers. Two transmission bands separated by a finite gap is observed with the given parameters. Note that, in contrast to the case in dielectric arrays where the bandgaps are determined from the edge of Brillouin zone, the bandgaps in plasmonic arrays are determined from the centre of Brillouin zone. This is a consequence of the inverted band curves in periodic plasmonic systems [11, 12]. Besides, as shown in Fig. 1(b), at some specific $t_m$ values, lower and upper band collides at the so-called Dirac point and the gap herein closes. Dirac points and their origins as well as their consequences on Bloch modes has been reported in Refs. [19, 20]. However, in these studies the systems were assumed lossless. In contrast, here we confirm the survival of Dirac point in the lossy cases. As expected, the magnitude of diffraction approaches infinity at the Dirac point. Thus, for the formation of localized mode at the vicinity of the Dirac point, a huge value of nonlinearity is required(to counteract the huge diffraction). Therefore, in the following we choose to work in the parametric spaces far away from the Dirac point. Several typical Bloch waves are shown in Fig. 1(c)-1(f) for transverse $E_x$ component of the field. As expected, modes at the base of each band ($k_x=0$) feature in-phase profiles at the adjacent lattices(unstaggered), while those at the edges($k_x=\pi/T$) feature out-of-phase profiles("staggered"). The $E_z$ components exhibit the same phase structures as $E_x$'s, but their amplitudes are continuous across the interfaces.

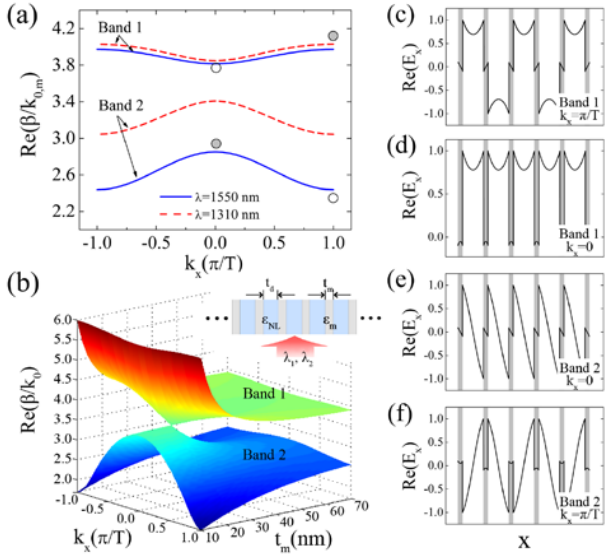

Fig. 1. (a) Bandgap structure of the plasmonic lattice. $t_m$=50 nm, $t_d$=240 nm. The solid (open) circles represent the location of self-focusing (defocusing) PLSs. (b) Bandgap spectrum vs. metallic layer thickness $t_m$, for $t_d$=240 nm, $\lambda$=1550 nm. The inset is a sketch of the present system. (c)-(f) Bloch waves ($E_x$). The gray (white) regions stand for metallic (dielectric) layers.

To search for the vector PLSs supported by the nonlinear plasmonic lattice, we write the field of the $m$th ($m$=1,2) components with the stationary form, $[E_{x,m}, H_{y,m}] = [u_m(x), v_m(x)]e^{i(\beta_m z - w_m t)}$ with $u_m$ and $v_m$ being modal profile and $\beta_m$ soliton propagation constant. Substituting them into MEs yields

$$\frac{k_{0,m}\varepsilon_{NL}}{z_0}u_m = \beta_m v_m \quad (1)$$

$$z_0\left[\frac{d}{dx'}(\frac{1}{\varepsilon_{NL}}\frac{d}{dx'})+1\right]v_m = \frac{\beta_m}{k_{0,m}}u_m \quad (2)$$

where the intensity-dependent dielectric permittivity $\varepsilon_{NL}$ is given by

$$\varepsilon_{NL} = n_{NL}^2 = [n_{linear,m} + n_2 \cdot \sum_m (|E_{x,m}^2| + |E_{z,m}^2|)]^2 \quad (3)$$

Here, $z_0 = \sqrt{\mu_0/\varepsilon_0}$, $x' = k_{0,m}x$ and $k_{0,m} = 2\pi/\lambda_m$.

Multiband vector PLSs are found by solving Eqs. (1) and (2) numerically with a self-consistent method. Figure 2(a) and 2(b) present one typical profiles of vector PLSs in self-focusing nonlinearity. The first component ($\lambda_1$=1310 nm) resides above the edge of its Band 1 and it is thus staggered in field profile. This component should be considered as nonlinear counterpart of Bloch wave shown in Fig. 1(c). The second component ($\lambda_2$=1550 nm) falls into the finite gap, being unstaggered as it has origin from the Bloch waves shown in Fig. 1(e). The intensity of the two components can be characterized by their associated maximum change of refractive index $\Delta n_1$ and $\Delta n_2$. The components hown in Fig. 2(a) and 2(b) have the same intensity $\Delta n_1 = \Delta n_2 =0.025$, requiring modal power of $P_1$=249 W/μm and $P_2$=264 W/μm, respectively. Here the power is defined as $P_m = (1/2)\int \text{Re}(E_{x,m}H_{y,m}^*)dx$.

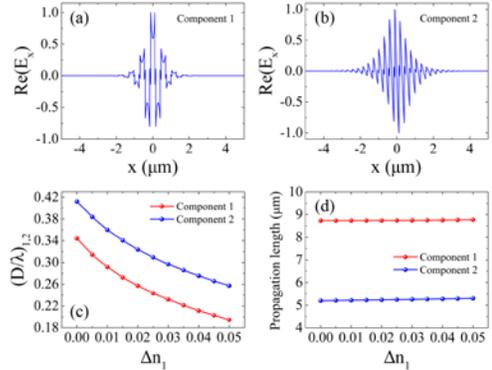

Fig. 2. (a), (b) Normalized electric field profiles of the self-focusing vector PLSs for $t_d = 240$ nm, $t_m = 50$ nm, $\Delta n_1 = \Delta n_2 =0.025$. $\lambda_1$=1310 nm, $\lambda_2$=1550 nm. (c) Soliton effective diameter vs. $\Delta n_1$, for $\Delta n_2$ =0.025. (d) Propagation length vs. $\Delta n_1$, for $\Delta n_2$ =0.025.

The particularly interesting feature of vector PLSs is that they provide a solution to trap and control multiple light signals in a very compact spatial dimension. Figure 2(c) plots the soliton effective diameter, $D_m = \sqrt{\int x^2 |E_m|^2 dx / \int |E_m|^2 dx}$, as a function of the nonlinear refractive index. The plot shows that deep-subwavelength confinement of both components is available, provided that refractive index change is on the order of $10^{-2}$. For example, for the soliton shown in Fig. 2(a) and 2(b), $D_1$=0.24 $\lambda_1$ and $D_2$=0.31 $\lambda_2$. They can be compressed further by increasing the intensity of either of the components. Interestingly, the propagation length of each mode, $\text{Im}(\varepsilon_m) \neq 0$, is only weakly affected by the intensity [Fig. 2(d)]. Obviously, for vector PLSs, only the shorter propagation length ($\min\{PL_m\}$) is practically meaningful, as that is the *effective interaction length* between the two components during their co-propagation. Therefore, the effective propagation length for vector solitons shown in Fig. 2(d) is min$\{PL_m\}$ =5 µm.

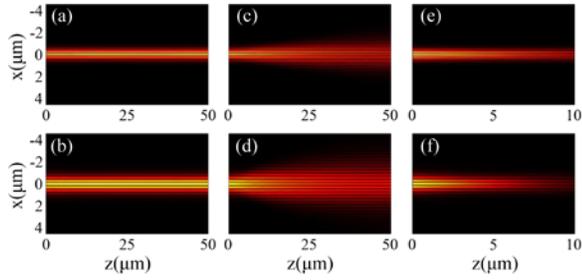

Fig. 3. (a), (b) Propagation of vector PLSs in lossless plasmonic lattices. (c), (d) Propagation of single component when the other is switched off. (e), (f) Propagation in real (lossy) plasmonic lattices. The first (second) row of panels corresponds to the Component 1 (Component 2).

To confirm that the above numerically found vector PLSs solutions are indeed stationary modes of the associated nonlinear plasmonic lattices, we feed the modes in Fig. 2 into the full nonlinear MEs and propagate them using a finite element method software (COMSOL Multiphysics). The results are shown in Fig. 3. It shows that both components nicely maintain their shapes over a long distance, if the metallic loss is ignored[Fig. 3(a) and 3(b)], or it is balanced by the background optical gain. In contrast, if only one component is present, discrete diffraction occurs [Fig. 3(c) and 3(d)]. This confirms that the formation of vector PLSs indeed relies on the mutually-supporting nature of the two components, and each component does not exist as solitons independently. The propagation of vector PLSs taking into account the realistic metallic losses is shown in Fig. 3(e) and 3(f), which, consistent with the mode analysis [Fig. 2(d)], show that the effective propagation length is determined by the propagation length of second component, i.e., min$\{PL_m\}$ = $PL_2$ =5 µm.

Vector PLSs are also found in self-defocusing plasmonic lattices. One representative profile is given in Fig. 4(a) and 4(b). Importantly, the effective propagation length of such mode is found as min$\{PL_m\}$ =16 µm. That is, compared with those in focusing nonlinearity, here PLSs could propagate over a distance essentially three times longer. The reason is that, in defocusing nonlinearity, both components oscillate in the same phase at the opposite sides of each metallic layers, as clearly illustrated in their associated Bloch waves [Fig. 1(d) and 1(f)]. In another word, both components are collective excitations of long-range SPPs and thus the vector states are of longer propagation length. In contrast, both components in vector PLSs in focusing nonlinearity are actually collective excitations of short-range SPPs and thus their effective propagation distance is rather limited.

In conclusion, we find the vector PLSs as two-component eigenmodes of the nonlinear Maxwell's equations. The two components are associated with different bands of transmission spectrum. We demonstrate their stationary and stable propagations, elucidate their subwavelength confinements as well as their distinct propagation lengths in self-focusing and self-defocusing nonlinearities.

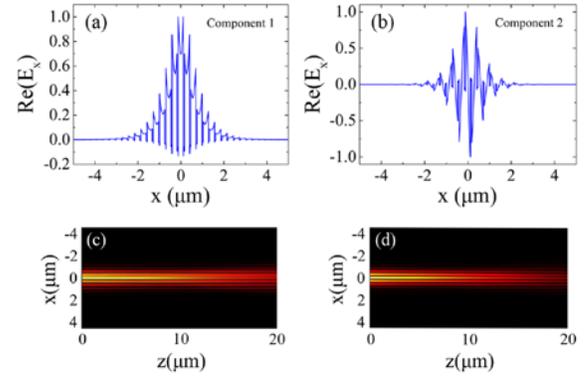

Fig. 4. (a), (b) Normalized electric field ($E_x$) profiles of the self-defocusing vector PLSs for $t_d$ = 240 nm, $t_m$ = 50 nm, $\Delta n_1 = \Delta n_2 = -0.025$, $\lambda_1$ =1310 nm, $\lambda_2$ =1550 nm. (c), (d) Propagation of the same modes in real (lossy) plasmonic lattices.

This research was supported by the NNSFC (11104181 and 61125503).